# Spatially evolving vortex-gas turbulent free shear layers: Part 2. Coherent structure dynamics in vorticity and concentration fields.

Saikishan Suryanarayanan[*] and Roddam Narasimha[+]
*Engineering Mechanics Unit, Jawaharlal Nehru Centre for Advanced Scientific Research, Jakkur, Bangalore – 560064, India*
Email : [*] saikishan.suryanarayanan@gmail.com;   [+] roddam@caos.iisc.ernet.in

This paper examines the mechanisms of coherent structure interactions in spatially evolving turbulent free shear layers at different values of the velocity ratio parameter $\lambda \equiv (U_1 - U_2)/(U_1 + U_2)$, where $U_1$ and $U_2$ ($\leq U_1$) are the free stream velocities on either side of the layer. The study employs the point-vortex (or vortex-gas) model presented in part I (Suryanarayanan & Narasimha, 2015, arXiv:1509.00603) which predicts spreading rates that are in the close neighborhood of results from most high Reynolds number experiments and 3D simulations. The present (2D) simulations show that the well-known steep-growth merger events among neighboring structures of nearly equal size (Brown & Roshko 1974) account for more than 70% of the overall growth at $\lambda < 0.63$. However the relative contribution of such 'hard merger' events decreases gradually with increasing $\lambda$, and accounts for only 27% of the total growth at the single-stream limit ($\lambda = 1$). It is shown that the rest of the contribution to layer growth is largely due to the increasing differential in size between neighboring structures as λ increases from 0.6 to 1.0, and takes place through two different routes. The first occurs via what may be called 'soft' mergers, involving extraction of little patches or filaments of vorticity over time from appreciably smaller (usually upstream) structures. This is consistent with earlier computational work on asymmetric two-vortex mergers (Yasuda & Flierl, 1995). The second route involves assimilation of disorganized 'vortex dust', which contains remnants of disrupted earlier structures not yet transferred to surviving ones. A combination of these processes is shown to cause an apparently nearly continuous increase in structure size with downstream distance at $\lambda = 1$, of the kind reported by D'Ovidio & Coats (2013) and attributed to mixing transition based on analysis of passive scalar concentration fields. However, simulations with tracer particles show that passive-scalar data are not reliable indicators of vortical structure dynamics, as they suggest dominance of continuous structure-growth even at smaller values of $\lambda$ (consistent with the simulations of Hernan & Jimenez (1982) and McMullan et al (2015)), in contradiction with the corresponding *vortical* structures growing via hard merger-dominated events. Thus structure- and layer-growth mechanisms observed in experiments and 3D LES in high Reynolds number shear layers are not inconsistent with explanations involving purely 2D mechanisms.

## I. INTRODUCTION

The turbulent mixing or free shear layer, the flow evolving downstream of a splitter plate separating two streams respectively with velocities $U_1$ and $U_2$ ($U_2 \leq U_1$ ; $\Delta U \equiv U_1 - U_2$, $U_m \equiv (U_1 + U_2)/2$), has been widely studied over more than six decades (from Liepmann & Laufer 1947 to McMullan et al 2015), as in some sense it is the simplest prototypical turbulent shear flow. Despite this extensive work, there are unresolved questions and controversies, central among which is the issue of universality, i.e. whether or not initial conditions are eventually forgotten by the layer to attain a unique self-preservation state with a universal growth rate.  Extensive simulations of a temporally evolving free shear layer by Suryanarayanan, Narasimha and Hari Dass (2013, henceforth SNH), via the vortex gas method, strongly suggested the existence of universality at least for this special case of 2D inviscid flow.  The vortex gas has been demonstrated to converge weakly to a smooth solution of the 2D Euler equations (Marchiro & Pulvirenti,1993). The observation that the large scale evolution is quasi-2D (Brown & Roshko, 1974, Wygnanski et al, 1979), the quantitative agreement of the temporal vortex-gas simulations (SNH) with experimentally observed initial transients of the harmonically forced mixing layer Oster & Wygnanski (1982), among other arguments (see SNH), suggest that the 2D inviscid results are indeed relevant for the 'real' (3D Navier-Stokes) plane turbulent free shear layers, although the issues concerning molecular mixing are outside the scope of the vortex-gas model.

Part I (Suryanarayanan & Narasimha, 2015, ArXiV : 1509.00603) of this paper extended the temporal study of SNH by presenting results for spatially evolving vortex-gas shear layers at different velocity ratios, with appropriate inflow and outflow boundary conditions. The 'equilibrium' / self-preservation spread-rates, so found in the zone independent of both inflow and outflow conditions, are broadly within the scatter of the experimental data. Furthermore, the results are in good agreement over a range of velocity ratios with most high-Reynolds number experiments beginning with Spencer & Jones (1971) to the more recent experiments (D'Ovidio and Coats, 2013, henceforth DC), as well as recent 3D Large Eddy Simulations (LES) ( McMullan, Gao and Coats, 2011, 2015, henceforth MGC1 and MGC2). This agreement was rather surprising, considering that DC, MGC1 and MGC2 conclude, based on analysis of their experimental and 3D LES results, that 2D models are inapplicable for studying even the large scale evolution of high Reynolds number shear layers. In this Part II, we address these issues.

Even though the evolution of the layer is basically statistical, what makes it particularly interesting is the visually dominating presence of large-scale ordered motion in the form of coherent vortical structures. Since their presence was demonstrated by the experiments of Brown and Roshko (1974), there has been substantial further work on the subject (see Brown & Roshko 2012 for a recent review). However, the interaction of the structures with each other, the role they play in the development of the layer, and whether (and if so how) they are affected by velocity ratio and small scale motion - these issues are not yet completely understood. More specifically, there has been the controversial question of whether the so-called 'mixing transition' (Brown & Roshko 1974, Konrad 1977, Dimotakis 2000), which greatly enhances small scale mixing and three dimensionality, has a significant effect on the large scale evolution of the layer (Brown & Roshko 2012, DC and MGC2).

Mergers between coherent structures have been widely accepted as the central growth mechanism of the free shear layer. This mechanism of growth was perhaps first proposed by Winant & Browand (1974, who use the term 'pairing') and Brown & Roshko (1974). However, there have been other views. Hernan & Jimenez (1982, henceforth HJ) analyzed cine-films of earlier experimental shadowgraph visualizations of a plane mixing layer with fluids of different refractive indices (Bernal, 1981). (The shadows depend on the second derivative of the index of refraction (Liepmann & Roshko, 1957), and therefore in this case correspond to a concentration field.) HJ concluded that most of the growth of the areas occupied by the coherent structures takes place between mergers rather than during merger events. They however carefully note that their results refer only to the evolution of the scalar field and do not necessarily apply to the behavior of the vorticity in the layer.

Recently an experimental investigation to understand the nature of the growth mechanism was carried out by DC. Their results, based on schlieren visualizations, suggest that there are two distinct mechanisms through which the coherent structures in the flow (and, by extension, the thickness of the layer) may grow. First is the mechanism of growth by mergers, as presented in Winant & Browand, 1974), where the sizes of the structures change little between well defined merger events (usually between two neighboring structures), but undergo a relatively sudden increase during such events. In the second mechanism each individual structure grows continuously and linearly with downstream distance, with mergers appearing to play an insignificant role in the overall evolution of the layer. DC concluded from their experiments that the flow evolution was dominated by merger processes before the onset of the 'mixing transition', and by the continuous growth of structures post mixing transition. In other words, the shift in growth mechanism to a continuous structure growth was attributed to the mixing transition and the associated 3D effects. These conclusions of DC were however drawn based on comparisons between pre- and post-mixing transition layers that did not have the same velocity ratio. For the constant density case (which will be the major concern of the present discussion), the pre-mixing transition layer (in which merger-dominated evolution was observed) had $\lambda = 0.627$ (where the velocity ratio parameter $\lambda \equiv (U_1 - U_2)/(U_1 + U_2)$), whereas the post-transition layer (in which continuous growth was observed) was a single-stream shear layer ($\lambda = 1$). Since the dependence of coherent structure dynamics on velocity ratio has not yet been studied in detail, one cannot rule out the possibility that the observed change in mechanism was a consequence of differences in velocity ratio.

In addition to experimental analyses of HJ and DC, continuous structure growth was most recently reported by MGC2 in their 3D LES study. But it is important to note that the visualization adopted in all these studies imply that they use passive scalar concentration fields to identify and track structures. It is well known that interpretation of passive scalar fields poses problems in turbulent flows, as they are not necessarily representative of the evolution of the underlying vorticity field (Babiano et al, 1987, Shraiman & Siggia, 2000). While there have been some recent studies (e.g. Gampert et al. 2014) that suggest that passive scalar fields are good candidates for identifying the turbulent-non turbulent interface, it is not clear whether such a conclusion would hold at high Reynolds number, and for identification of coherent structures or other such contexts.

In the light of the above issues, a more detailed understanding of coherent structure dynamics is essential. Therefore, following the study in Part I of this paper, we here examine the instantaneous evolution of the vortex locations in the spatially evolving vortex-gas free shear layer to understand how coherent structures interact and impact the growth of the layer. We perform this analysis at different values of $\lambda$ to investigate whether such an interaction mechanism changes with velocity ratio. Further, we also study passive scalar fields via tracer particles.

The organization of this paper is as follows. In section II, we briefly describe the problem setup. Section III presents the evolution of the coherent structures at different values of $\lambda$, including 0.627 and 1.0, the two velocity ratios employed by DC. Section IV presents a detailed analysis on the dynamics of vortical coherent structures at $\lambda = 1$ and discusses the underlying (2D) mechanisms. In Section V, we compare evolution of passive scalar fields at $\lambda = 0$ (temporal), 0.627 with the evolution of the respective vorticity fields. In Section VI, we discuss our results in relation to the observations and conclusions of earlier work, before concluding in Section VII.

## II. PROBLEM SETUP

The computational setup used for the spatial shear layer simulations presented here is shown in Fig.1A and described in detail in Part I. It incorporates certain improvements over that presented by Basu et al (1992, 1995), and can be briefly described as follows. The computational domain of $L \times \infty$ initially contains $N_0$ point vortices of equal strength $\gamma = L(\Delta U)/N_0$ with average $x$-spacing $l \equiv L/N_0$. There is a single semi-infinite vortex-sheet upstream of the computational domain to mimic the splitter plate and a fan of semi-infinite vortex sheets to represent the far-field downstream condition. The strength of the downstream vortex-sheets is such that it generates a Gaussian vorticity profile, and the angle of spread of the fan is set to match the spread rate of the layer computed within the domain.

The vortices are released into the computational domain at the edge of the splitter plate at a constant frequency in accordance with the Kutta condition, and the given *y*-displacements drawn from a uniform random distribution of amplitude *a*. It was shown in part I that, especially towards the single-stream limit ($\lambda \to 1$) where the fluctuation in the number of the vortices in the domain can be large, it is essential to introduce a buffer vortex. This buffer vortex is located at a fixed position downstream of the computational domain but has a temporally varying strength that ensures conservation of global circulation. The velocity induced at the point vortices is computed at each time-step as a sum of the contribution of the Biot-Savart interactions with other vortices, the upstream and downstream vortex sheets and the buffer-vortex, and their positions are updated using an RK-4 scheme with a time-step $\Delta t = 0.1 l/U_m$ ($= 0.1 (2\lambda) l/\Delta U$). Note that the Hamiltonian was conserved to within $10^{-5}$ of its initial value in temporal simulations (SNH) with $\Delta t = 0.1 l/\Delta U$. Furthermore, increasing or decreasing the time-step by a factor of two is observed to change evolution of thickness by less than 2%. More details are presented in Part I.

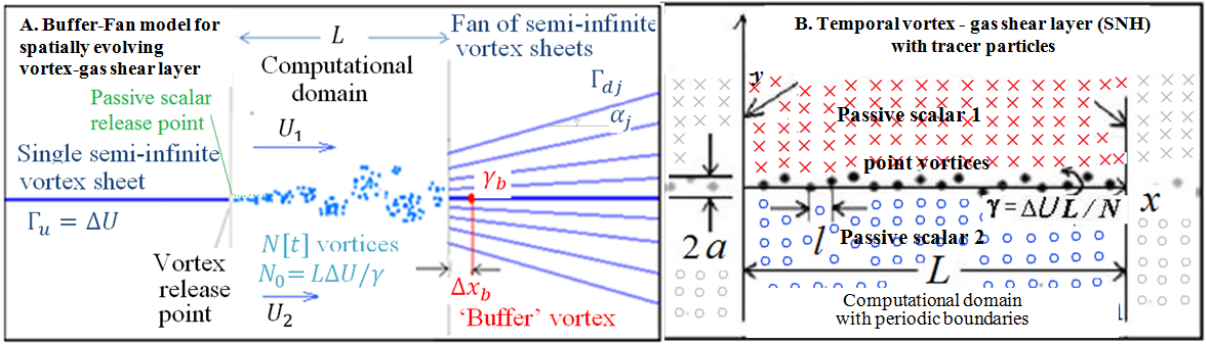

**Figure 1.A.** The present buffer-fan model for spatially evolving vortex-gas shear layer. Passive scalar field is generated by tracer particles that are injected on the high-speed side, $4l$ above the vortex-release point. **B.** Temporally evolving vortex-gas shear layer (SNH) with two different kinds of tracer particles above and below the centre-line.

For studying the evolution of passive scalar fields, we make use of non-diffusive tracer particles which are essentially advected by the velocity field. When the flow becomes statistically steady (typically after a few convective times ($L/U_m$)), neutrally buoyant tracer particles are released on the high-speed side at the same frequency as the vortices, at a normal distance $4l$ above the splitter plate. This mimics introduction of a dye or smoke filament on the high speed side in an experimental flow visualization. In addition to the $N_0 = 2000$ simulations presented in part I and analyzed here in Section III, we also present high resolution simulations with 8000 vortices and 8000 tracer particles in Section V. In order to reduce the computational costs, the vortices and downstream fan in these simulations are initialized using data from an instantaneous snapshot from the statistically steady state part of the $N_0 = 2000$ simulations. (However each vortex is replaced by 4 vortices in its neighborhood, each with a fourth of the strength of the original vortex). This way the time taken for the flow to reach steady state is drastically reduced, and analysis can start from the next convective time. We shall also present analysis of coherent structures from vorticity and passive scalar fields of a temporal free shear layer (shown in Fig.1B), using the method described in Sec V and SNH.

# III. COHERENT-STRUCTURE DYNAMICS AT DIFFERENT VELOCITY RATIOS

## A. Evolution of coherent structure trajectories and sizes for $\lambda$ = 0.627 and 1.0

In this section, we analyze the evolution of coherent structures in vortex-gas shear layers at different velocity ratios ($N_0 = 2000$, $a/l = 0.001$, buffer-fan setup, see Part I), each at its respective self-preserving flow state (Regime II, see part I for details) independent of initial and boundary conditions. We first report two illustrative cases of vortex-gas shear layers at $\lambda = 0.627$ and $1.0$, precisely the same velocity ratios as used by DC in their constant-density experiments.

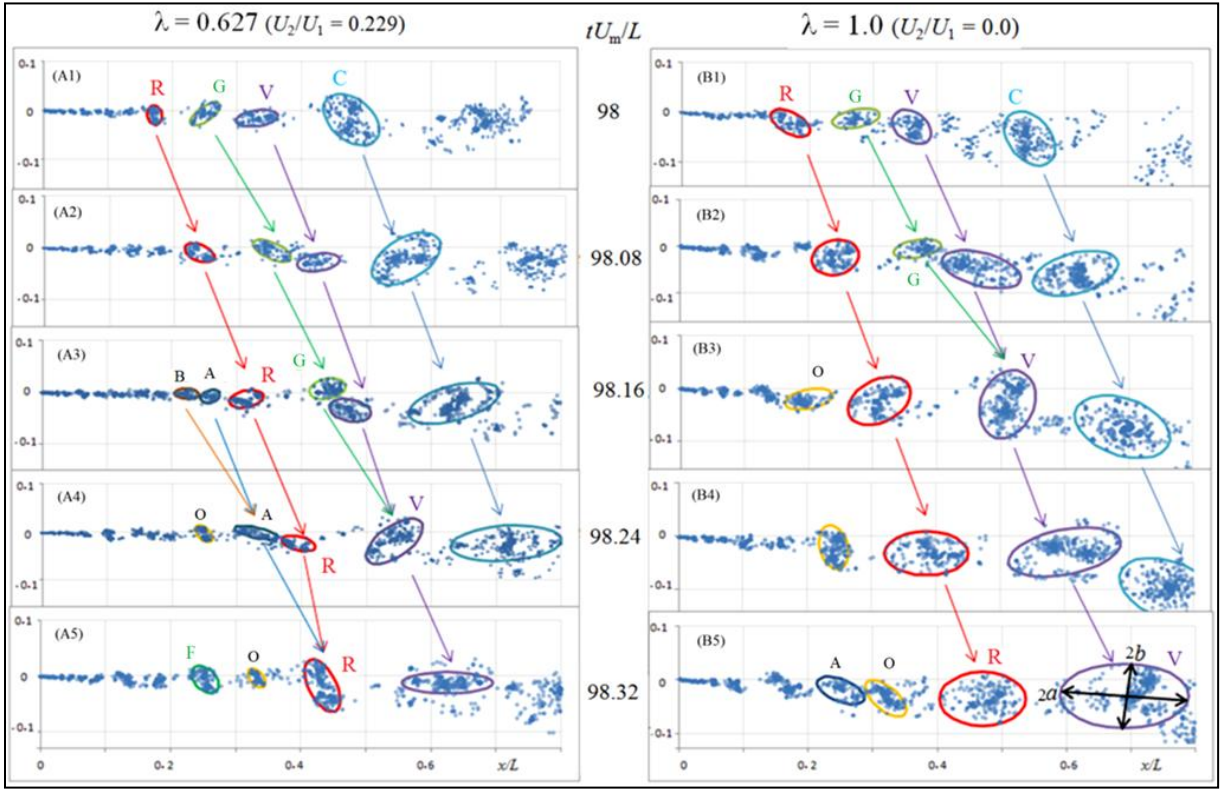

**Figure 2.** Sample comparison of evolution of structures at $\lambda$ = 0.627 (A1-A5) and 1.0 (B1-B5) in the present vortex-gas simulations

The evolution of the vortex locations in the two cases are shown in Fig. 2A and B respectively. At both velocity ratios the vortices can be observed to clump to form coherent structures that increase in scale with downstream distance. There are often one or more relatively small high vorticity core regions around which the rest of the structure (of a relatively lower vorticity) is organized. There are also much sparser braid regions that connect neighboring structures. The braids are not considered as a part of the structure, by the convention adopted here, unless they have been entwined around what is identified as a single structure, when, they are considered as the part of the coherent structure. The coherent structures can be well described by visually fitting ellipses, also shown Fig.2. From the major ($a$) and minor ($b$) axes of the ellipse, an effective radius of the structure can be defined as the geometric mean, $r_{ab} = \sqrt{ab}$. (It has to be noted that some earlier studies such as DC fit circles, while HJ fit ellipses. The latter seems to be a more visually faithful choice and is therefore adopted here.) The following convention is used in the naming and tracking of structures. Structures at the specified initial instant are designated with a single capital letter (that is often chosen to represent the starting letter of the color of the ellipse used to mark them). The structure that results from the merger of two (or more) structures takes the label of the originally farthest downstream structure. The evolution of the trajectories and sizes of the various structures identified in Fig.2A and B (and also for later times not shown in Fig.2) are tracked in Figs.3A and B respectively.

Figures 3(A1) and 3(B1) show the *x-t* trajectories of the coherent structures for the two velocity ratios. It can be observed that the trajectories are broadly similar and that there are mergers in both cases. However, the difference between the two velocity ratios is seen in Figs. 3(A2,A3) and 3(B2,B3), which plot the size (i.e. $r_{ab}$) of the structures against their *x*-location.

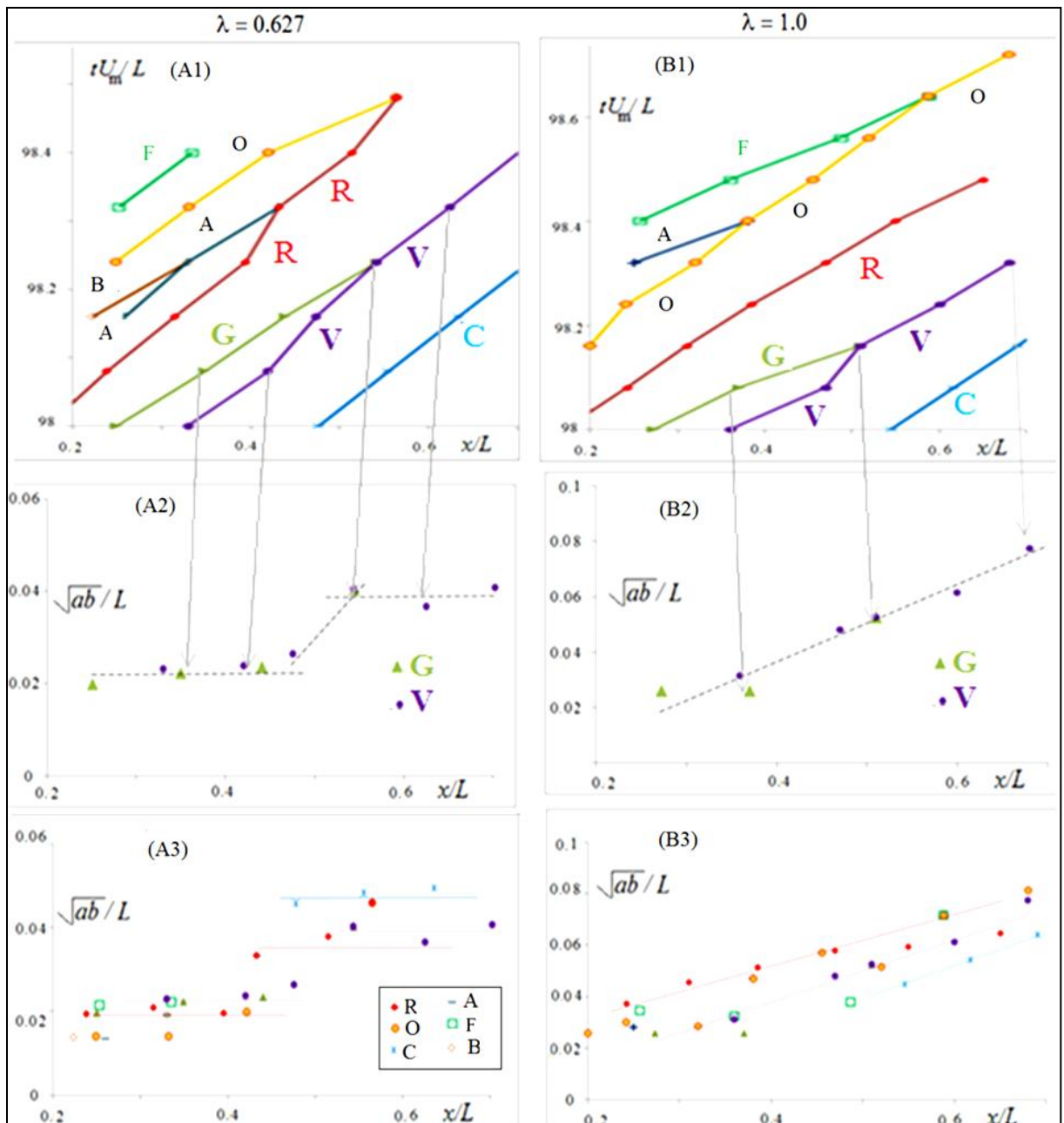

**Figure 3.** Comparative evolution of downstream location and size of structures at $\lambda$ = 0.627 and 1.0.

Consider the evolution of the structures labeled G and V for $\lambda$ = 0.627 shown in Fig.2A and 3A2. Clearly, the size of either structure hardly changes (< 20%) between Figs.2A1-A2 (over *x/L* of 0.25 to 0.475). On the other hand the well-defined merger in Fig.2A3-A4 leads to a new structure (again labeled V) that has an effective radius that is 154% that of the larger of the original structures and over 80% of the sum of the effective radii of the original two structures. Therefore this is a steep-growth event which we shall term as a 'hard merger' (we shall introduce the concept of 'soft mergers' - another kind of interaction between two structures at a later stage). Further, as seen in Fig.3A2, the new structure after the merger hardly changes in size (<1%) as it moves from *x/L* = 0.54 to 0.7. Similar observations of structures undergoing steep-growth during mergers and hardly experiencing change in size in-between mergers can be made for several other structures for $\lambda$ = 0.627 in Figs.2A and 3A3, following the familiar scenario described in Section I. We shall use the term 'merger-dominated' to describe this mechanism of layer growth in the rest of the paper. (Aside: For $\lambda$ = 0.627, it is observed that during mergers between any two structures 1 and 2, $(r_{ab})_{\text{new}} \sim (r_{ab})_1 + (r_{ab})_2$ on an average, Thus the area, $\pi r_{ab}^2$, of the new structure is larger than the combined areas of the original structures ($\pi(r_{ab1}^2 + r_{ab2}^2)$) before merger. This implies that the average vorticity within the new structure is lower, and this is consistent with equilibrium similarity theory as the vorticity of a structure scales with the local mean vorticity which decreases with downstream distance.)

The above-described scenario of merger-dominated evolution for $\lambda$ = 0.627 may be contrasted with the results at $\lambda$ = 1 shown in Figs. 2B and 3B, where structure R, for example, does not appear to undergo any distinct hard merger event (expect for one with a very small unlabelled structure at $tU_m/L$ = 98.16), yet grows to nearly twice its original size as it travels downstream from *x/L* ~ 0.2 to ~ 0.5, from Fig.2B1 to Fig.2B5. The structures labeled G and V do undergo a hard merger just before $tU_m/L$ = 98.16 (Fig.2B3), but overall evolution of V does not seem to be greatly affected by the merger. As a consequence, if one follows the larger structure V alone from $tU_m/L$ = 98.0 to 98.32 (Fig.2B1 to B5), it appears to grow continuously and almost linearly by 246% with time and downstream distance. (An examination at an intermediate time of 98.12 $L/U_m$ (not shown here) suggests that the merger indeed causes a small step increase in size of V, but it contributes to just 3% of the increase in size of V from $tU_m/L$ = 98.0 to 98.32 and is thus barely noticeable.) Similar observations can be made for several other structures, and we shall discuss in detail the mechanisms, including soft mergers, by which structures grow between hard mergers in Sec. IV.

It is however important to note that hard mergers are to an extent responsible for the growth of the layer at even the higher velocity ratio. One way of quantifying the role of hard mergers in the overall growth of the layer is by the ratio of the spatial rate of growth in structure size in-between mergers (which is taken as average change in $r_{ab}$ of structures between mergers divided by the average *x*-distance travelled by structures between mergers), to the average value of $r_{ab}/x$. It is found that this value, which represents the contribution of growth between hard mergers to the overall growth of structures (and hence the layer), increases from 28% for $\lambda = 0.627$ to 73% for $\lambda = 1.0$. While there is a certain degree of subjectivity involved in fitting the ellipses as it is done visually, this is a limitation shared with several other studies including that of DC, and does not affect the main conclusion. Also the number of structures used in the present analysis, again comparable to that in the DC analysis, leads to a statistical uncertainty of the order of 10%, but this is small compared to the difference between the results for the two cases, which vary by a factor of over 2.5. This analysis was also repeated over another randomly picked time interval ($tU_m/L$ = 82 to 82.5), and led to similar results (see Fig.A1 in Appendix). Qualitatively similar results are obtained for other downstream boundary conditions (Fig. A2 in Appendix), indicating the robustness of the conclusion.

**B. Mid-region circulation distribution for $\lambda$ = 0.627 and 1.0**

However the above analyses cover relatively short times of order $L/U_m$ (as in DC analysis), so in order to obtain more representative statistics over longer timescales, we perform the following independent objective analysis. We compute, at each of the several time instants separated by $tU_m / L=0.008$ over the period $tU_m / L = 20$ to $80$, the number of vortices (i.e. the total circulation) within a region $\mathcal{R}$ defined as $\{L/2 − δ_ω(L/2)\} < x < \{L/2 + δ_ω(L/2)\}$, $-\infty < y < \infty$ (see Fig. 4C). The average effective diameter of the structure is $O(δ_ω)$ ($λ = 0.627 : 2r_{ab} \sim 0.13x, δ_ω \sim 0.1x$; $λ = 1.0 : 2r_{ab} \sim 0.22x, δ_ω \sim 0.185x$). The average separation between structures is therefore over twice the effective diameter, so the sampling region $\mathcal{R}$ contains a single structure on an average (Fig. 4C). We compute the local maxima of the circulation over a duration $2U_m/ δ_ω(L/2)$, which corresponds to the average structure passage time. Hence this statistic is expected to reflect the average circulation of structures passing through $x = L/2$. Note that there is no subjectivity involved in this analysis other than the selection of the box size, which is chosen to be $2δ_ω(L/2)$ for both cases, i.e. the same fraction of the respective local thickness. The probability distribution function (PDF) of the local maxima of circulation is computed from the time-series for $λ = 0.627$ and $1.0$ simulations, and shown in Fig. 4.

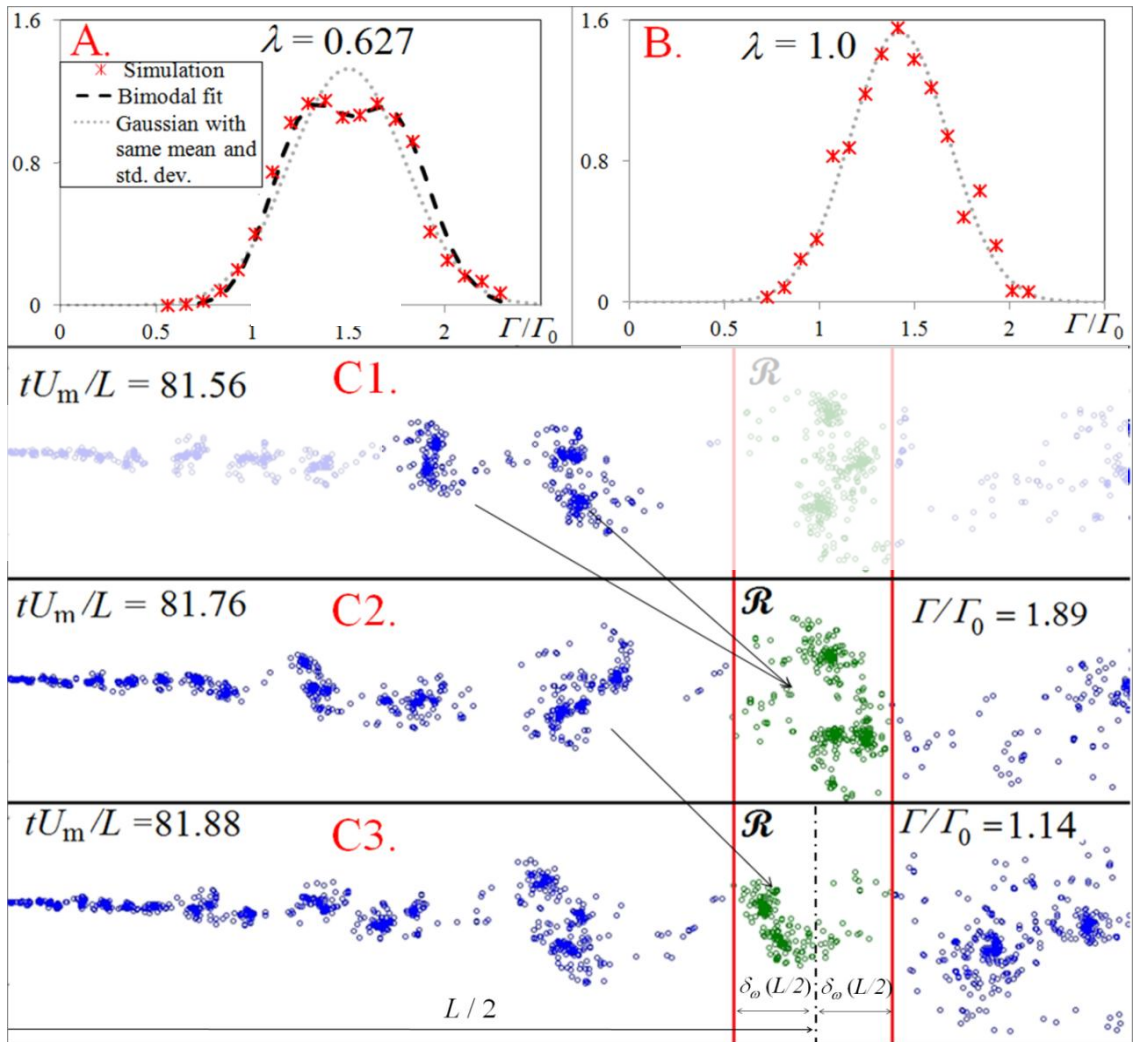

**Figure 4.** Comparison of PDF of local maxima of circulation in the region $\mathcal{R}$ (see text) at A. $λ = 0.627$ and B. $λ = 1.0$. The abscissa is normalized by $Γ_0 = 2δ_ω ΔU$. Note that the distribution at $λ = 0.627$ agrees with a bi-modal fit as opposed to the Gaussian corresponding to the mean and variance. This is in contrast with the distribution for $λ = 1$, which is adequately described by a uni-modal Gaussian. C. Snapshots explaining the origin of bimodality at $λ = 0.627$.

## C. Results at other values of $\lambda$

For $\lambda = 0.627$ the distribution of circulation happens to be bimodal, and is well represented by a sum of two displaced Gaussians ($=A_1 G_1(\mu_1, \sigma_1) + A_2 G_2(\mu_2, \sigma_2)$) with $A_1 = 0.47$, $A_2 = 0.53$, $\mu_1 = 1.275$, $\mu_2 = 1.72$, $\sigma_1 = 0.185$, $\sigma_2 = 0.203$; where $G_1$ and $G_2$ are Gaussian distributions with means $\mu_1$ and $\mu_2$ and standard deviations $\sigma_1$ and $\sigma_2$), indicating that the structures that pass $x = L/2$ have two preferred values of strength. An examination of the vortex snapshots shown in Fig.4C provides an explanation for this bimodal distribution. Fig. 4C(1-2) show that when two structures merge just before entering $\mathcal{R}$, the resulting steep-growth causes a high value of circulation in $\mathcal{R}$ contributing to the higher value mode in the PDF. On the other hand, as seen in Fig. 4C(2-3), when such a merger has not taken place just before $\mathcal{R}$, it leads to an observation of a significantly lower value of circulation, in the range of the mode with the lower value in the PDF. This, therefore, is consistent with the earlier observation of structure size strongly affected by mergers, with the two preferred size ranges corresponding to structures that have undergone mergers immediately upstream and downstream of $\mathcal{R}$. On the other hand, we observe that the PDF at $\lambda = 1.0$ is well described by a standard (uni-modal) Gaussian, supporting the conclusion that structure growth in the single-stream case is not dominated by steep-growth merger events.

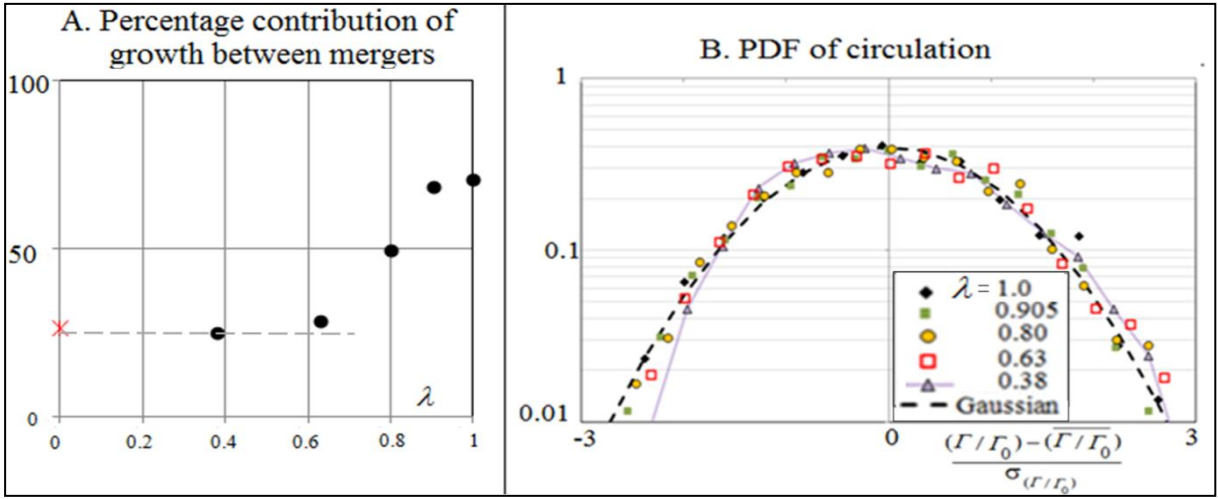

**Figure 5.** Effect of velocity ratio on **A.** Percentage contribution of continuous structure-growth between mergers to overall growth of structures. The results from analysis of temporal simulations (SNH) are shown at $\lambda = 0$. **B.** PDF of circulation of local maxima of circulation in the region $\mathcal{R}$. The distributions are normalized to have zero mean and unit standard deviation.

The above results, taken together, establish that there are two distinct mechanisms of structure growth at $\lambda = 0.627$ and $1.0$. We now examine the behavior at other values of $\lambda$. We find (from Fig.5A) that, for the temporal layer (SNH, also see Fig 10A in Sec V) which represents the $\lambda \to 0$ limit, as well as for $\lambda = 0.38$, continuous growth between structures does occur, but contributes to less than 30% of the net growth of structures. Taken with the result at $\lambda = 0.627$ (Figs.2A, 3A and 4A) suggests that the evolution is merger-dominated at least upto $\lambda = 0.627$. Of interest are intermediate points between $\lambda = 0.627$ and $1.0$, which can throw light on the nature of the transition from the merger-dominated to continuous-growth dominated mechanisms. For this purpose, we make simulations at $\lambda = 0.8$ and $0.905$ (with $N_0=2000$, $a/l = 0.001$). We find (Fig.5A) that, at $\lambda = 0.8$, both mergers and continuous growth play nearly equal roles, and that the continuous structure-growth mechanism is dominant at $\lambda = 0.905$, contributing to about 70% of the total growth (also see Fig.A3 in Appendix). Figure 5B extends the results of the analysis presented in Fig.4 for a range of velocity ratios. As noted in Fig.4, the PDF for the $\lambda = 1$ case is well described by a (unimodal) Gaussian distribution, while the highest deviation from the Gaussian is observed for $\lambda = 0.38$. The results for the intermediate values of $\lambda$ lie between these two extreme values. Thus, these results taken together strongly suggest that the growth mechanism smoothly shifts from a merger-dominated to a continuous structure-growth dominated mechanism with increase in $\lambda$, and that the behavior at $\lambda = 1$ is neither singular, nor the unique peculiarity of the single-stream case.

In summary, we find that there is a gradual transition in the growth mechanism from merger-dominated to continuous-structure-growth dominated regime with increase in $\lambda$. Thus (hard) mergers between coherent structures continue to be present at all values of velocity ratio, but their role as steep-growth events and their relative contribution to the overall increase in the size of the structures and the layer growth become less significant with increase in $\lambda$, being less than 30% at $\lambda = 1$. Since the present simulations are purely 2D, explanations for the continuous-growth mechanism based on mixing transition (as suggested by DC) are not relevant here; explanation by a purely 2D mechanism must be possible. We explore this issue in the following section.

## IV. 2D MECHANISMS FOR CONTINUOUS STRUCTURE- GROWTH

To sketch here a plausible physical explanation for the observed change in the layer growth mechanism as $\lambda$ increases towards unity, consider first the detailed flow evolution shown in Fig. 2B for $\lambda = 1$. Figure 6 tracks every vortex that belonged each of the four individual structures at $tU_m/L = 98.0$, over the time span $97.84 \leq tU_m/L \leq 98.32$. It can be seen that the increase in size of the structure V (for example), from $tU_m/L = 98.0$ to $98.16$ has the following three contributions.

(i) The hard merger with G completed just before $tU_m/L = 98.16$, as shown in Fig. 6E.

(ii) Capture of vortices from the structure G (see Fig. 6D) by the structure V before the actual merger of the two structures. The interaction between the structures R, O and A at $tU_m/L = 98.32$ (Fig. 7G) exhibits the same pattern. Therefore this process can be called a soft-merger event, which is defined as an interaction between neighboring structures involving vorticity transfer but without changing the total number of structures.

(iii) The capture of the unorganized background 'vortex dust' marked 'D' in Fig.6 (colored black) and present between the structures V and C by $tU_m/L = 98.0$ (Fig. 7C). This collection of vortices has a circulation per unit area that is less than a fourth of that of the neighboring structures. Thus it does not constitute a coherent structure, and hence is not marked with an ellipse in Fig. 2B1. By tracking the history of these unorganized vortices, it is observed that their lifetime as a collection is less than a fourth that of other structures. The origin of some of the vortex dust can be traced back to a merger between two structures (between $tU_m/L = 97.84$ and $97.92$ in Fig.6A-B) that eventually results in V. The rest of the dust is torn from the edges of the structure C. The above argument of interaction of a structure with vortex dust also applies for unusually small structures.

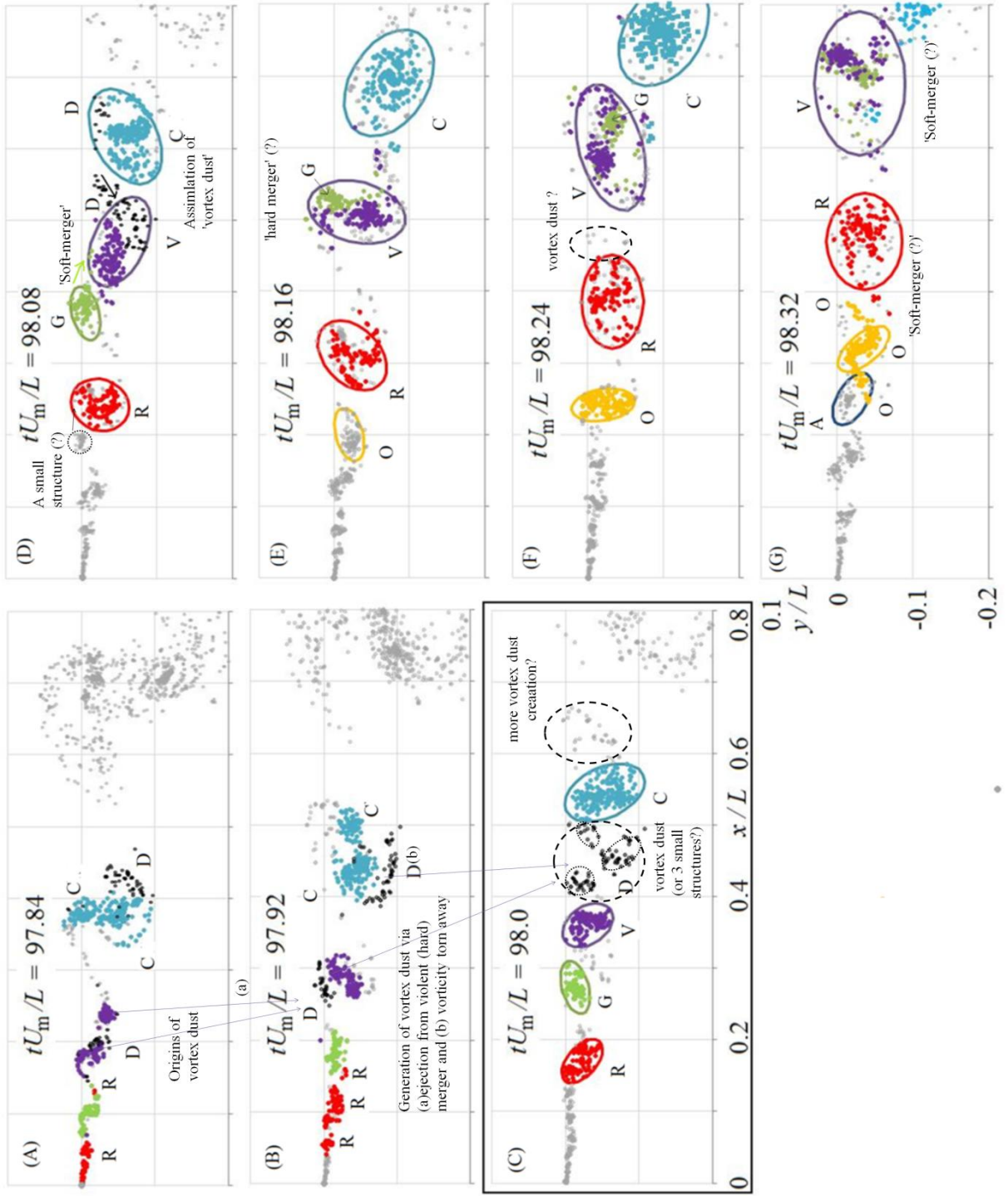

**Figure 6.** History and evolution of vortices belonging to individual structures for $\lambda = 1$ at $tU_m/L = 98.0$, except for vortices of structure O, that are tracked from $tU_m/L = 98.24$. The vortex dust D (shown in black), which lies between the structures V and C at $tU_m/L = 98.0$ is not part of any organized vortex structure, are tracked over $97.84 \leq tU_m/L \leq 98.08$. Vortices that are not tracked are shown in gray.

This scenario can be summarized as follows. First of all, the disparity in size between adjacent structures in the layer is related to the layer spread-rate, which in turn increases with $\lambda$ (see part I), as the average structure size at any *x*-location scales with the local layer thickness. From Fig.3B, we estimate an average of 41% difference in size between merging structures just before the three identified merger events at $\lambda = 1$. This can be contrasted with the $\lambda = 0.627$ case in Fig.3A, where the average disparity in the size of the participating structures is 26% in four merger events, out of which three cases involve disparity of less than 13%. This suggests the following scenario. The large size-disparity between interacting neighboring structures as $\lambda$ increases towards unity is responsible for the observed continuous structure growth dominance, via soft mergers - the complex interactions of the larger downstream structure feeding on the vorticity of either the smaller neighboring structure upstream, or on the disorganized background vorticity, with the latter being ejected during previous violent mergers or torn from outer parts of larger structures during interactions with neighboring structures. (In practice, objectively distinguishing between small structures, filaments of smaller upstream structures and disorganized vorticity is difficult, and therefore we do not attempt the analysis of their relative contribution statistics to the observed continuous-structure growth.)

The idea of the soft merger is indeed consistent with what is well known from asymmetric two-vortex merger studies (e.g. Yasuda & Flierl, 1995) and can be illustrated via the cartoon shown in Fig. 7. The left panel of Fig.7 shows an idealized interaction between two representative structures at small values of $\lambda$, which are usually of comparable size and strength. In such cases, the two structures tend to go around each other and undergo a distinct hard merger over a short time, resulting in a new highly elliptical structure that is much larger than either of the original structures (such as the V-G merger in Fig.2A). On the other hand, when one vortex is much larger or stronger than the other, as is usually the case as $\lambda \to 1$ shown in the right panel of Fig.7, we observe a soft merging process during which the former (such as structure V in Fig.6) tends to filament ( or shred) the smaller and/or weaker (usually upstream) vortex (such as G) into thin strands and absorbs the vorticity (by entwining the strands). This increases the size of the larger and stronger vortex more or less continuously in time (or equivalently streamwise location) till the small structure is completely absorbed, often with a hard merger at the end (consistent with the evidence presented in Fig 6 D-E). The weaker vortex can either be an upstream structure (such as the structure G in Fig.6) or disorganized vortex dust (such as D in Fig.6). Hard mergers do take place, but since they involve a dissimilarly sized structures, the relative increase in the size of the larger structure is smaller than that involving similar structures (see G-V interaction Fig.6 D-E). Furthermore, such hard mergers often are violent interactions (unlike hard mergers for smaller values of $\lambda$) that result in ejection of smaller structures or vortex dust (as seen in Fig.6A-C, thereby both further reducing the steep-increase in size of the resulting structure as well as creating opportunity for future continuous growth. Thus, as $\lambda \to 1$, when the evolution of individual structures is tracked in space (such as in Figs. 2B, 3B), many of them tend to exhibit apparently continuous and approximately linear growth between merger events. This can last over a substantial part of lifetime of the structure (e.g the size of structure R is observed to increase nearly linearly over 66% of the time it spends in the self-preservation regime of the domain). This description shows that whether the layer growth is merger-dominated or continuous structure-growth dominated depends on the size disparity between neighboring structures, which in turn depends on the velocity ratio through the layer spread rate.

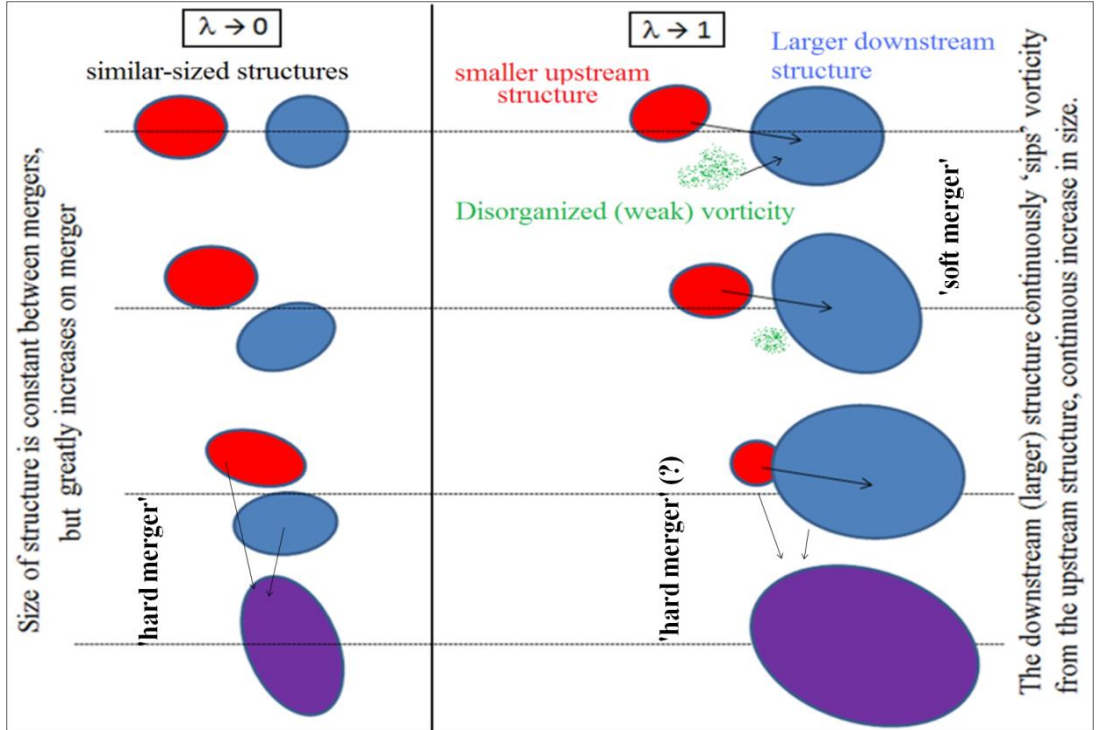

**Figure 7.** Cartoon illustrating the proposed role of dissimilarity in structure size on evolution.

## V. EVIDENCE FROM PASSIVE SCALAR FIELDS

In plane mixing layers with 3D turbulence, the evolution of the passive scalar field is different from that of the vorticity field owing to (i) the presence of the vortex-stretching term $(\boldsymbol{\omega}\cdot\nabla)\mathbf{u}$ in the vorticity equation (and the absence of an equivalent in the passive scalar evolution), and (ii) the possibly different coefficients of kinematic viscosity and scalar diffusivity. These two factors will not be discussed in the present paper, where we shall confine our attention to the relationship between the scalar and vorticity fields for the idealized 2D inviscid case with zero scalar diffusivity. Even in this case, the relationship is non-obvious even though the vorticity and passive scalar concentration are governed by the same advection equation (other than the Biot-Savart coupling of the vorticity and velocity fields). This is because the passive scalar concentration field at the initial instant or location is not always identical to the initial vorticity field, and this flow is chaotic. While dye, smoke film and (for fluids with different refractive indices) optical shadowgraph techniques have been widely used to obtain a qualitative picture of the flow field, including the presence of coherent structures, we shall examine the difficulty of using such information for deeper analysis of the vorticity field in relation to vortex-gas shear layers. We shall consider the following three cases in subsections VA, VB and VC respectively. The first is a spatial shear layer with $\lambda = 0.627$, in which tracer particles are injected just above the splitter plate on the high speed side. The second and third involve temporal shear layers in which the top and bottom fluids are initialized with two distinct kinds of tracer particles. While we shall focus on the self-preservation regime in the first two cases, the initial conditions for the final case are chosen to reveal the dynamics in the regime preceding self-preservation.

**A. Spatial shear layer (λ = 0.627) at self-preservation, with scalar particles introduced just above splitter plate.**

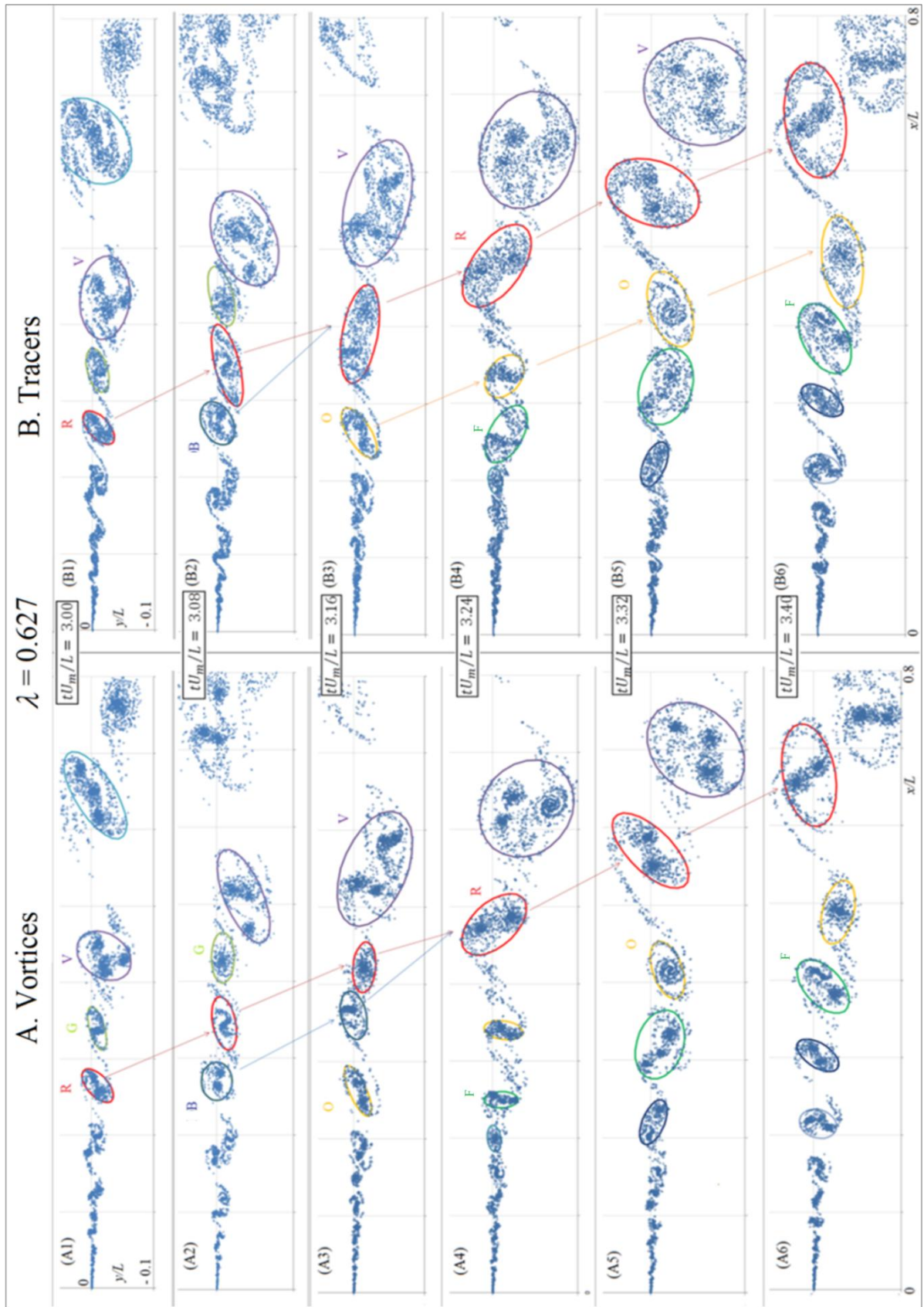

**Figure 8.** Evolution of (A) vortex locations and (B) tracer particles for $\lambda = 0.627$

For the first case, we perform high resolution simulations with $N_0 = 8000$ and with 8000 tracer particles ($N_s$) for $\lambda = 0.627$, with initial conditions computed from $N_0 = 2000$ simulations (see section II for details). The latter are injected into the flow at $x = 0, y = 4l$. Figure 8A shows the evolution of the vortex positions (representative of a continuous vorticity field in the limit $N_0 \to \infty$), and 8B shows the corresponding locations of the tracer particles (representative of a continuous passive scalar field in the limit $N_s \to \infty$).

The dynamics of the vortical structures shown in Fig. 8A are consistent with that depicted in Fig. 2A for $N_0 = 2000$. The results are not only qualitatively similar but vary in terms of average structure size by less than 10% (within the statistical uncertainty of the short time of simulation) over the four-fold increase in number of vortices. Most significantly, the vortical structures grow via the merger-dominated mechanism for both values of $N_0$, with continuous growth contributing to only 25% in the high resolution simulation (only slightly different from the 28% for $N_0 = 2000$, the difference being within the statistical uncertainty associated with the small number (<10) of structures tracked in each case). Thus these results provide mutual validation, suggesting the adequacy of 2000 vortices used earlier as well as that of the boot-strap initial condition and short integration time used in the high resolution simulations presented in this section.

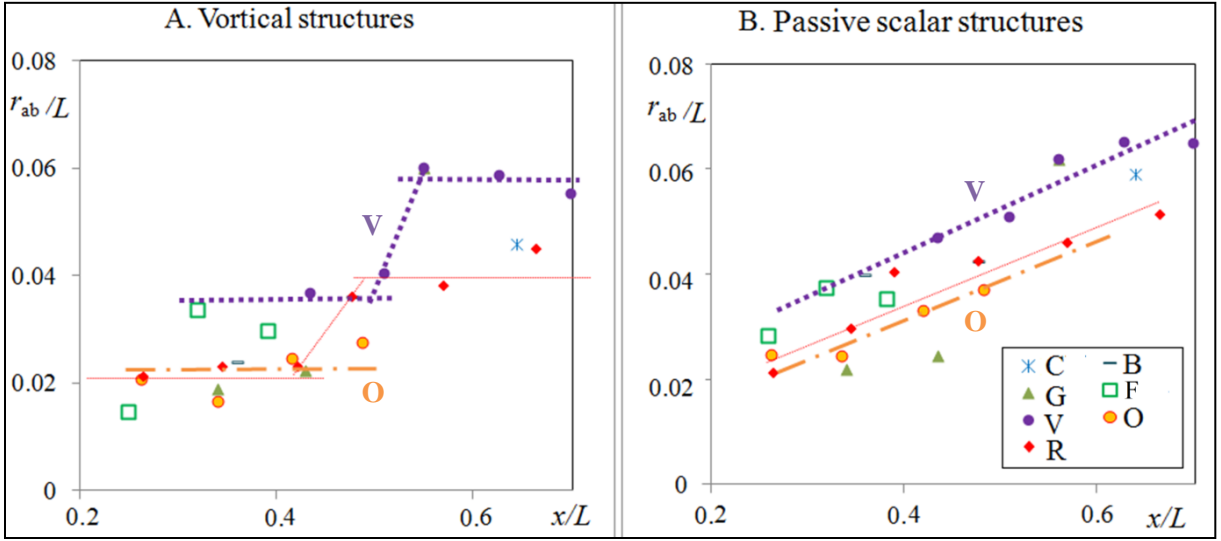

**Figure 9.** Evolution of (visually) identified structure-size with downstream location for (A) vortex structures and (B) passive scalar structures in the self-preservation regime of a spatial vortex gas shear layer at $\lambda = 0.627$.

While the passive scalar fields may appear broadly similar to the respective vortex distributions at first glance, a more careful examination of Fig.8 reveals the following specific differences in the evolution of the two fields. (i) The passive scalar structures are less distinct and on an average larger (by about 20%). (ii) Considering the structure V in the passive scalar field, it is clearly seen to rotate counter-clockwise from $tU_m/L = 3.00 - 3.08$ (in Fig. 8B1-B2), in contrast with the respective vortex structure V, suggesting orientation dynamics of passive scalar structures do not always follow that of the respective vortex structures. (When the high speed side is on top, the vortex-structures always rotate clockwise, and this can be seen for all structures in Fig 8B that are beyond $x/L \sim 0.4$, but not for passive scalar structures as noted above; the frequency of the snapshots presented in Fig.8 is not adequate to track the rotation of the smaller structures further upstream). (iii) Figure 8A4 shows that the vortical structures R and B merge at $tU_m/L = 3.24$, whereas their passive scalar analogs complete merging earlier at about $tU_m/L = 3.16$ (as seen in Fig.8B3). Thus passive scalar fields can lead to an incorrect identification of the structure boundary as well as a different conclusion on the time at which a merger event occurs, in part because the 'braids' observed in the passive scalar fields are much thicker than those in the vorticity field and in part because the edges of the structures, which have low vortex-concentration, often still have moderate passive scalar concentration (as observed for structure V in Fig.8). In summary, the above observations broadly suggest that vorticity and passive scalar concentration fields are not quite the same. This is not surprising if we recognize the motion of vortices (even at the level of coherent structures) is inherently chaotic, and the associated chaotic vorticity field can advect the passive scalar particles away from the vortices over time.

We now address the main question of evolution of the structure size with downstream distance, when inferred from the passive scalar field. Figures 9A and B plot the effective radius of the identified structures in the vorticity and passive scalar fields respectively, with downstream location (as in Fig3A3 and B3). Consider at first the evolution of the structures V and O. In the vortex picture (Fig.9A), V experiences a steep growth during the merger with G and remains of constant size before and after the merger. On the other hand the passive scalar analog (Fig.9B) of V exhibits continuous growth. The domination of continuous growth in-between mergers can be observed in several other passive scalar structures in contrast their respective vortex structures which exhibit merger-dominated growth. Overall, analysis of data in Fig.9 shows that continuous growth in the passive scalar structures contributes to nearly 50% of the net growth of the structures, and this is twice the value (25%) observed in the vorticity field.

Thus, these simulations suggest that while the interaction dynamics of vortical structures transition to domination (over 50% contribution) of continuous structure-growth only for $\lambda \gtrsim 0.8$, the passive scalar structures exhibit a similar proportion of continuous growth at the lower value of $\lambda = 0.627$, at which the associated vortex structures are undergoing merger-dominated growth. To explore this issue further, we now consider the temporally evolving free shear layer (SNH) which represents the $\lambda \to 0$ limit. As noted in Sec.III (Fig.5), the vortical structures in the temporal limit are observed to undergo merger dominated growth.

## B. Self-preservation dynamics of vorticity and scalar concentration in temporal shear layers ($\lambda = 0$)

We consider a realization of the temporal case RI in SNH, integrated upto $t\Delta U/l = 1800$ from an initially equispaced-in-$x$ configuration of 3200 vortices with $y$-positions generated from a uniform random disturbance of amplitude $a = 0.05l$ about the $x$-axis (more details in SNH). The irrotational fluid on each side is uniformly filled with 46400 neutrally-buoyant non-diffusive tracer particles over upto $|y|/l = 232$, which extends to nearly twice the maximum extent attained by the vortical layer in the simulations. The tracer particles in the top and bottom fluids are tracked separately, and their evolution in the limit of a large number of particles represents the concentration field of the two fluids initially present at the top and bottom of the vorticity layer. This initial condition corresponds to that in experimental flow visualization of fluids with different refractive indices (such as DC and HJ) or numerical simulations where the mixture fraction of the two fluids is used for analysis (e.g. MGC2).

A sample evolution for the vorticity and scalar fields in this flow is shown in Fig.10 A and B respectively. We use the outermost interface lines (defined as separation between top fluid and mixed fluid, or mixed fluid and bottom fluid, or top fluid and bottom fluid) as a guide to fit ellipses in the passive scalar field, consistent with the approach of HJ. We then repeat the analysis carried out in VA for temporal free shear layer considered here.

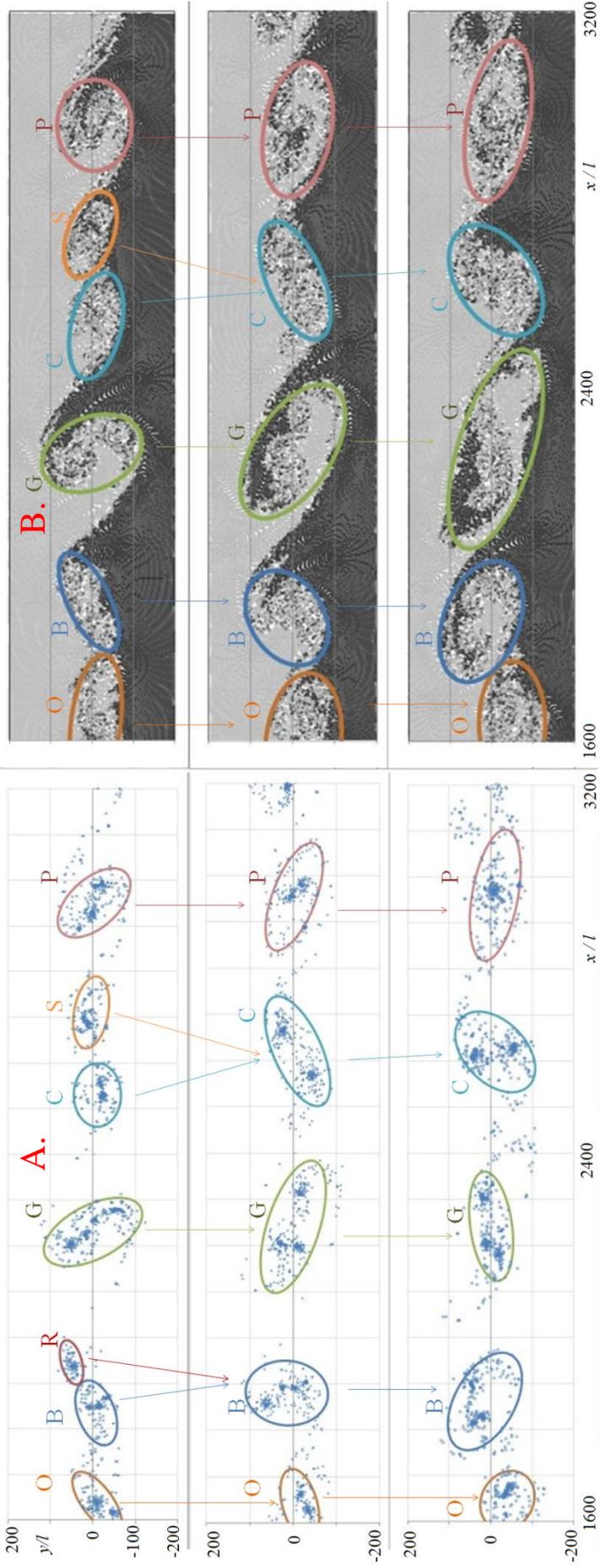

**Figure 10.** Comparative evolution of vorticity and passive scalar fields in temporal shear layer for $t\Delta U/l = 1200$ (top panels), 1400 (middle) and 1600 (bottom panels)

A possible explanation for the observed continuous growth of the passive scalar structures at the lower values of $\lambda$ (0.627 and 0) follows from the argument presented in Sec. V for continuous growth of vortical structures as $\lambda \to 1$. In this turbulent/chaotic flow, some passive scalars eventually occupy high vorticity regions such as the core of a structure, while others continue to be present in the low vorticity boundaries and the braids between the structures. As discussed in Sec. IV, the high vorticity regions tend to stretch, filament and eventually wind the low vorticity regions around them. When the vorticity field is examined, the distinction between the high and low vorticity regions is clear. Thus a structure identified by a high vorticity region grows (at $\lambda \lesssim 0.8$) relatively less by this mechanism and increases in size predominantly when interacting closely with another high vorticity region. This happens via a steep-growth event of a well defined merger. On the other hand, the passive scalar field, while loosely following the vortical structures, does not always distinguish between high and low vorticity regions. Thus structures identified by the passive scalar field grow between merger events (which are much less distinct), by the entwining regions with low (or even zero) vorticity but with moderate concentration of passive scalars.

## C. Pre-self-preservation dynamics of vorticity and scalar concentration in temporal shear layers ($\lambda = 0$)

We now consider the third case, which is also a temporal shear layer with the same number of vortices and tracer particles as in the second case, but with *a/l* = 16. This higher value of initial amplitude of vortex *y*-displacement leads to an initially higher vorticity thickness. It was observed that the corresponding initial instability wavelength is also larger, leading to a longer relaxation time to self-preservation ($t\Delta U/l \sim 1000$ in contrast with $t\Delta U/l \sim 20$ in the previous case), and therefore permits a detailed study of the evolution of (coherent structures in) vorticity and passive scalar fields before attaining self-preservation. Focusing our attention on the time internal $500 \leq t\Delta U/l \leq 900$, it can be observed from Fig.11 that during this regime coherent structures are present. However, it was inferred from the evolution of thickness (not shown here) that the flow has not attained self-preservation during this stage of the evolution. This non-universal pre-self-preservation coherent structure-dominated regime is labeled RIb (following SNH, Suryanarayanan, Brown & Narasimha (under preparation)), with R Ia indicating the initial evolution preceding the formation of coherent structures.

It can be observed from Fig 11, that there is no major qualitative difference in the evolution of the coherent structures observed in the vorticity picture between Regime Ib ($500 \leq t\Delta U/l \leq 900$ in Fig.11) and Regime II (Fig.10, $t\Delta U/l \geq 1000$ in Fig.11), in the sense that the evolution is dominated by merger-driven steep-growth events in both regimes. The contribution of continuous growth was found to be 30.8% during $500 \leq t\Delta U/l \leq 900$ (Regime Ib) and 29.7% for the evolution $1000 \leq t\Delta U/l \leq 1800$, neither very different from the self-preservation value of 26.3% observed in Sec VB.

We now examine the right panel in Figure 11. First consider the latter phase of the evolution, $1000 \leq t\Delta U/l \leq 1800$, which corresponds to self-preservation or near self-preservation in the growth of thickness. It can be seen that in this regime the coherent structures, as identified from the passive scalar concentration field, display continuous growth. While mergers do take place, the overall growth is dominated by continuous increase in structure size in-between such merger events. We find contribution of growth between mergers to the overall growth of the structures is 68.1 %, a value close to the 70.3% observed for the case discussed in Sec. VB. This result is therefore consistent with observations made so far, namely that coherent structure dynamics in the passive scalar picture is dominated by continuous structure-growth, though the respective vortical structures follow merger-dominated growth.

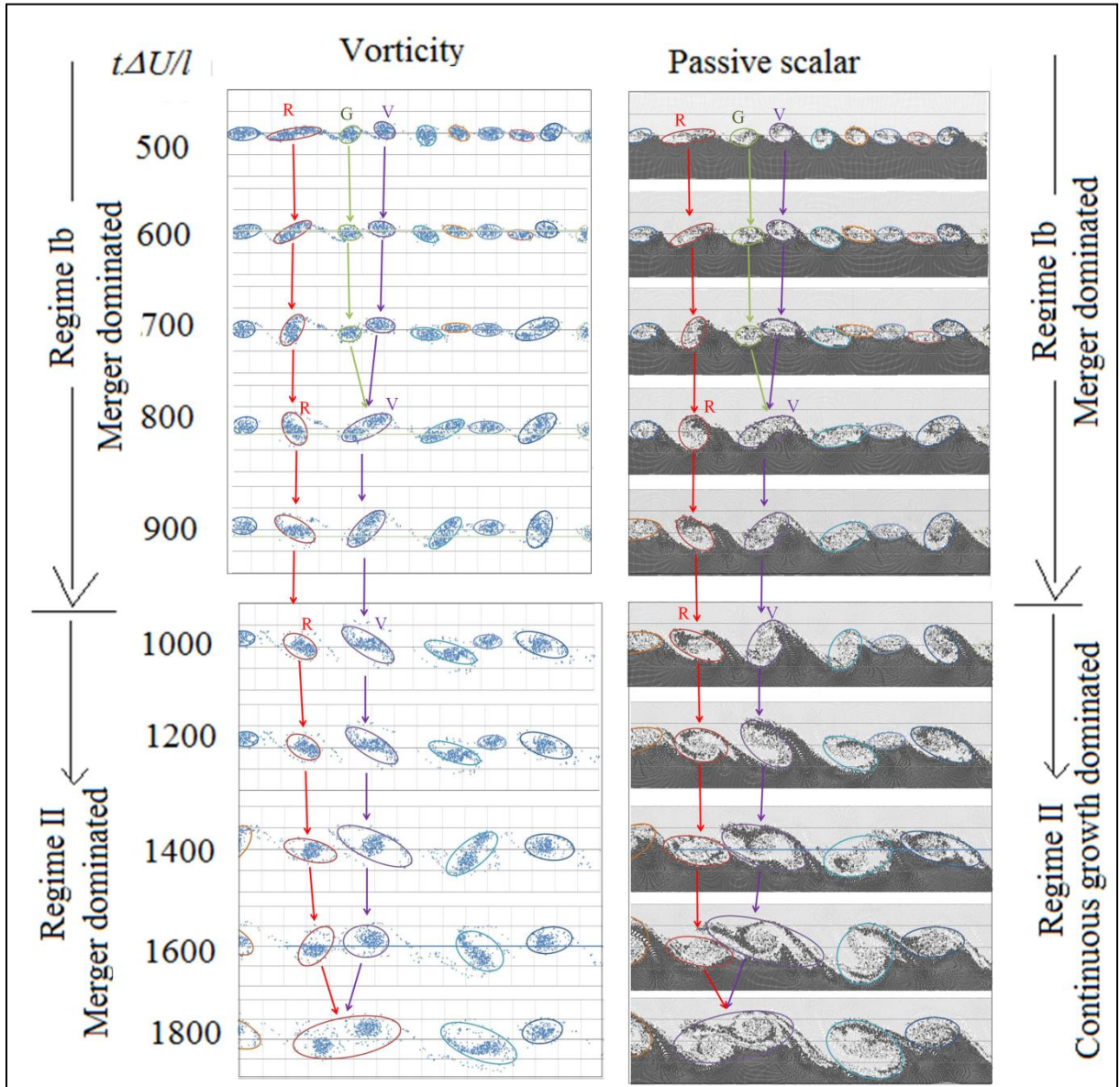

**Figure 11.** Coherent structures dynamics as inferred from vorticity (left panel) and passive scalar concentration fields for temporal vortex gas in Regimes Ib (pre-self-preservation) and II (self-preservation). The evolution of the structures R, G and illustrate that while growth of both vortical and scalar structures is dominated by mergers in RIb, the growth of scalar structures become continuous-growth dominated on approaching RII.

Interestingly however, a different picture emerges in the early evolution of the coherent structures in Regime Ib that precedes self-preservation. It can be seen from Fig.11 that, for $500 \leq t\Delta U/l \leq 900$, both vortical structures and passive scalar structures grow predominantly via steep-growth merger events, with continuous growth in-between mergers contributing to only 30.8% and 32.5% respectively. A possible explanation for this behavior is that mergers are essential in causing the passive scalar field to become distributed sufficiently widely (aided by chaos), and thus to depart from the dynamics of the associated vorticity field. (In the example considered with random initial conditions, it takes at least one set of merges and this duration also coincides with the relaxation to self-preservation. Depending on the specific initial condition, there is no reason why this process cannot take more mergers.)

## VI. IMPLICATIONS OF PRESENT RESULTS

### A. Effects of velocity ratio vs. mixing transition

With regard to the effect of velocity ratio on the growth mechanism of coherent structures and on the relation between vorticity and passive scalar fields, the present results are relevant from a fundamental point of view and can be expected to directly translate to planar free-shear layers before the onset of mixing-transition. However, as it was shown in SNH and Part I, the spread rates predicted by the present simulations closely agree with data from available high Reynolds number experiments (such as Spencer & Jones, 1971, $Re \sim 10^5$; DC, $Re$ : upto $4 \times 10^4$) and recent 3D LES simulations (MGC2, $Re$: $3 \times 10^4$). The Reynolds number in all the above examples clearly exceeds the mixing transition Reynolds number ($\sim 10^4$); the occurrence of mixing-transition was confirmed in DC and MGC2 via span-wise measurements. Therefore, the agreement of the vortex-gas spread-rate with the above experiments and simulations suggests that a 2D description is fully capable of describing the large scale evolution of even post-mixing-transition planar free shear layers.

The above inference is in contradiction with the conclusion of DC that there is a change in the fundamental mechanism of layer growth with mixing-transition, based on their observation of merger-dominated evolution in a pre-mixing transition shear layer and continuous-structure-growth in a post-mixing transition shear layer. The present work suggests that a plausible explanation for this observation is that DC did not account for the effect of the velocity ratio. In the constant-density experiments of DC, the pre-mixing-transition layer, where merger-dominated evolution was observed, had $\lambda = 0.63$. The post-mixing-transition layer, where continuous structure-growth was observed, had $\lambda = 1.0$. Thus, velocity ratio was another parameter that was different between the two experiments. While the results of the present 2D simulations cannot probe into the effects of mixing transition, they certainly throw new light on the effect of velocity ratio (in the 2D case). For the same two velocity ratios used by DC, we find the same two mechanisms of structure growth. (It must be noted that the DC results show lower scatter in the size vs. $x$ diagrams than the present vortex-gas simulations. This could be due to the span-wise averaging the DC analysis implies. Though the large scale is quasi-2D, the differences between the different 2D sections could have contributed to smoothen the statistical differences in individual structure size). We also explain the 2D mechanisms that result in this apparently continuous-structure growth in our purely 2D simulations, namely soft mergers and generation and assimilation of vortex dust. This shows that the differences in mechanism noted by DC across the two constant density experiments can be adequately explained by differences in the velocity ratio, without appealing to mixing transition or associated 3D mechanisms.

Based only on the present simulations we cannot directly comment on the non-uniform density experiments, but it has to be noted that the above observation that the change in mechanism with increase in spread rate, as elaborated in Sec. IV, is consistent with some results in the non-uniform density experiments of DC. For example, the spread rate at $\lambda = 0.627$, with $\rho_2/\rho_1 = 7.2$, is about 1.5 times (DC, Brown & Roshko 1974 for results at density ratio of 7) than that in the uniform density case at the same velocity ratio, and is in fact close to the spread rate for $\lambda = 0.9$, $\rho_2/\rho_1 = 1$. The observation of continuous growth by DC for this case is thus consistent with the present explanation.

## B. Interpretation of flow visualizations that exploit scalar concentration fields

However, it is important to note that the DC and HJ visualizations imply measurement of scalar concentration fields. While the general differences between the vorticity and scalar concentration fields in turbulence are well known, we present an analysis specific to evolution of coherent structures (in the 2D case) via the present vortex-gas simulations using tracer particles (Sec. V). We note that the passive scalar field can be misleading as indicators of the vorticity field evolution, suggesting incorrect orientation dynamics, as is clear from an examination of Fig. 10 of HJ for several structures. More significantly, we show that, even though the growth of the vortical structures is clearly dominated by steep-growth merger events (with continuous growth contributing only 26.3%) for $\lambda \to 0$, the evolution of structures as identified in the concentration field is dominated by continuous-structure-growth (70.3%) in the self-preservation regime. This evidence is further supported by analysis of the structure spacing shown in Table.1.

One way to measure structure spacing is the streamwise distance between the centers of the visually fitted ellipses. Another method is to measure the separation between successive zero-crossings of the 'braid' as adopted by Bernal (1981). Both of these methods are shown in Fig.12. While the braid-crossing method has the advantage that the crossings are usually well defined and is therefore more objective, the method counts two structures whose braid is entirely located below or above the centerline as one, as shown in Fig. 12 . Therefore it is not surprising that the braid-crossing technique, results in (upto 20%) higher mean spacing ($\mu_l$) than centre to centre method. However, there is no noticeable difference in $\sigma_l / \mu_l$.

**Table.1.** Comparison of mean ($\mu_l$) and standard deviation ($\sigma_l$) of coherent structure spacings in the vorticity and passive scalar fields in the present simulations with experimental visualizations.

|  | Technique | Field | $\mu_l/\delta_\omega$ | $\sigma_l/\delta_\omega$ | $\sigma_l/\mu_l$ |
|---|---|---|---|---|---|
| Brown & Roshko ($U_2/U_1 = 0.38$, $\rho_2/\rho_1 = 7$) | Centre(?) | Passive scalar | 2.9 | 0.93 | 0.32 |
| Bernal ($U_2/U_1 = 0.38$, $\rho_2/\rho_1 = 1$) | Braid | Passive scalar | **3** | 0.72 | **0.24** |
| Temporal vortex gas $N = 32000$ | Braid | Vorticity | 3.17 | 1.26 | 0.40 |
| Temporal vortex gas $N = 3200$ | Centre | Vorticity | 2.49 | 0.86 | 0.35 |
| Temporal vortex gas $N = 3200$ | Braid | Vorticity | 2.91 | 1.20 | 0.41 |
| Temporal vortex gas $N = 3200$, $N_s = 89600$ | Centre | Passive scalar | 2.67 | 0.74 | 0.27 |
| Temporal vortex gas $N = 3200$, $N_s = 89600$ | Braid | Passive scalar | **3.21** | 0.88 | **0.27** |

We also find from Table.1 that the passive scalar field shows 10% larger mean, and more significantly 30% lower standard deviation. Thus $\sigma_l / \mu_l$ for the vorticity can be 50% higher than what is measured from the corresponding vorticity field. This is consistent with the argument of continuous-growth domination (and closer to unimodal distribution) of passive scalar structures vs. the steep-growth domination (and more strongly bimodal distribution as in Fig.4A, with a larger standard deviation) of the corresponding vortical structures. The values of $\mu_l/\delta_\omega$ and $\sigma_l/\mu_l$ obtained from applying the braid technique on the passive scalar fields in the present (temporal) simulations are within 15% of Bernal results, suggesting that the overall dynamics of coherent structure in the scalar field indeed mimic that in real post-mixing-transition shear layers.

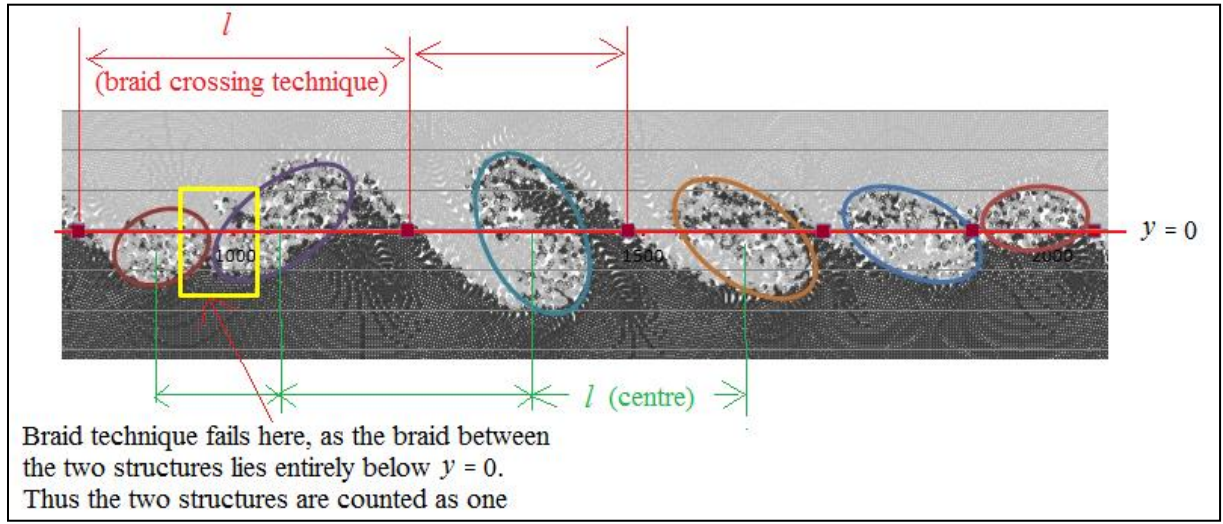

**Figure.12.** Methods of measurement of inter-structure spacing.

Thus the observations of continuous growth of structures reported for constant-density cases at lower values of $\lambda$ by the experimental visualizations at $\lambda = 0.45$ by HJ and the MGC2 3D LES at $\lambda = 0.61$ are consistent with the present study of passive scalar fields.

**Table.2.** Summary of results on effect of velocity ratio on coherent structure dynamics in vorticity and passive scalar fields. M indicates merger dominated structure growth, C indicates continuous growth domination.

| Velocity ratio | Vorticity field | | Concentration field | |
|---|---|---|---|---|
| | RIb | RII | RIb | RII |
| $\lambda \to 0$ | M | M | M | C |
| | | | M (DC) | C (HJ) |
| $\lambda \to 1$ | | C | | C(DC) |

However, this picture, appears, at first sight, not capable of explaining the merger dominated growth observed by DC at $\lambda = 0.63$. This is where it becomes important to recognize that the above conclusions only apply to self-preserving shear layer dynamics and that, as noted by DC, the layer in the $\lambda = 0.63$ experiment has not attained self-preservation. Analysis of temporal vortex-gas free shear layers, shows that while the scalar field dynamics differ from the vorticity dynamics in the self-preservation regime, both show merger dominated growth in the regime that precedes self-preservation. Thus the present results are indeed consistent with the data of DC, HJ and MGC2, and this is summarized in Table.2.

# VIII CONCLUDING REMARKS

In part I, we reported agreement of the present 2D vortex-gas simulations with high Reynolds number experiments and 3D simulations over a range of velocity ratios. This suggests that vorticity dispersal by purely 2D mechanism provides relatively satisfactory prediction of the large scale dynamics of a turbulent free shear layer even post mixing transition. Here in part II, the analysis of the evolution of structure sizes and locations is presented.

It has been suggested that there are two distinct mechanisms by which a turbulent free shear layer may grow. The first is well known, and involves a rapid increase in size of the coherent structures (and hence the layer thickness) during a hard merger of two (occasionally more) neighboring structures. The second involves a gradual but nearly continuous growth of each structure, with hard mergers playing an insignificant role. This has been proposed (DC, MGC2) as the dominant growth mechanism for the layer after mixing transition, which occurs at Re $\sim 10^4$. The purely 2D inviscid simulations reported here, using the vortex-gas technique, show that layer growth occurs principally through steep-growth hard merger events for velocity ratio parameter $\lambda \lesssim 0.7$. However the role of the continuous-structure growth which accounts for less than 30% of the overall growth for $\lambda \leq 0.627$, increases thereafter, accounting for half of the total growth at $\lambda = 0.8$ and rising to 73% by $\lambda = 1$. Deeper analyses reveal that the origins of the apparently continuous growth of individual structures lie in the soft merger process characteristic of interaction between dissimilarly sized structures as shown in earlier two-vortex merger studies (Yasuda & Flierl, 1995), and a cycle of creation and subsequent assimilation of disorganized vorticity (or small structures).

Further simulations of the flow initialized with tracer particles, as in typical experimental visualizations or in computer simulations of passive scalar fields, reproduce passive scalar structures that grow via mergers in the regime before self-preservation, but exhibit continuous structure growth in the self-preservation regime. This observation is significantly different from the respective vortical structures at the same values of $\lambda (= 0.627$ and $0)$, which grow via hard-merger-dominated mechanism both before and during the self-preservation regime. These results demonstrate that, contrary to what has been recently suggested (DC), continuous-structure growth is entirely possible through purely 2D mechanisms, not dependent on the 3D motions accompanying the mixing transition. Furthermore, depending on the value of $\lambda$, anywhere from 30 to 70% of the layer growth is due to rapid merger events. The present work thus suggests that conclusions of experiments at two different values of $\lambda$ (0.627 and 1.0, the latter after mixing transition, DC), has to be interpreted with caution, as the effect of higher $\lambda$ can be easily misattributed to effect of the transition. Correspondingly, a purely 2D model does not prove that a mixing transition cannot have an additional effect on the growth mechanism beyond that of the velocity ratio, although we are aware of no evidence of a difference in overall growth rate of the layer thickness before and after the mixing transition occurs in a given flow. Further work is needed to separate the effect of the two parameters namely $\lambda$ and Re.

# ACKNOWLEDGEMENTS

We thank Prof. Garry Brown (Princeton) and Prof. Anatol Roshko (Caltech) for many rewarding and enjoyable discussions and suggestions. We acknowledge support from DRDO through the project RN/DRDO/4124 and from Intel through project RN/INTEL/ 4288.

## APPENDIX

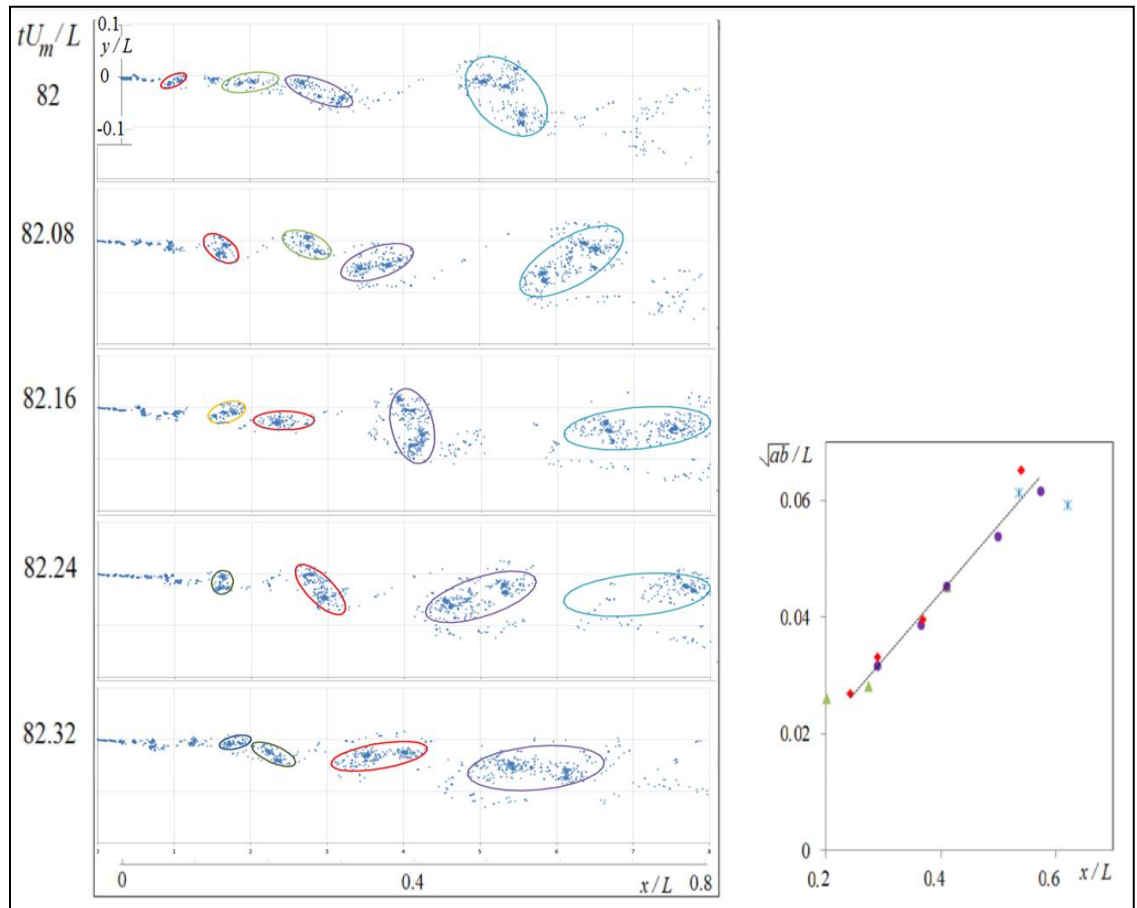

**Figure A1.** Analysis for $\lambda = 1$, shown in Figs. 7 & 8 for $tU_m/L = 98$ to 98.4 repeated for $tU_m/L = 82$ to 82.4. Note the qualitatively similar results are observed, suggesting the robustness of the conclusion on continuous growth of structures at $\lambda = 1$.

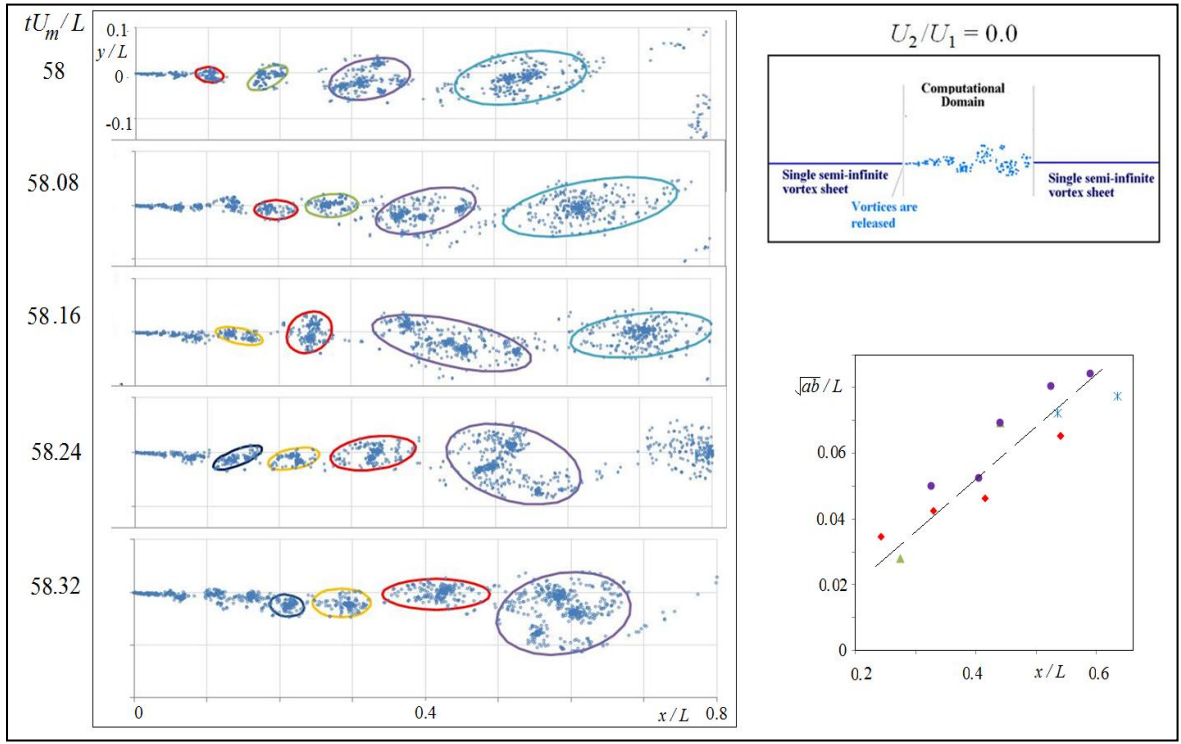

**Figure A2.** Analysis for λ = 1, shown in Figs. 2 & 3 for buffer-fan model repeated for a setup with a single downstream vortex-sheet and without doublet. The results suggest the robustness of the conclusion on continuous growth of structures at λ = 1 to different downstream boundary conditions.

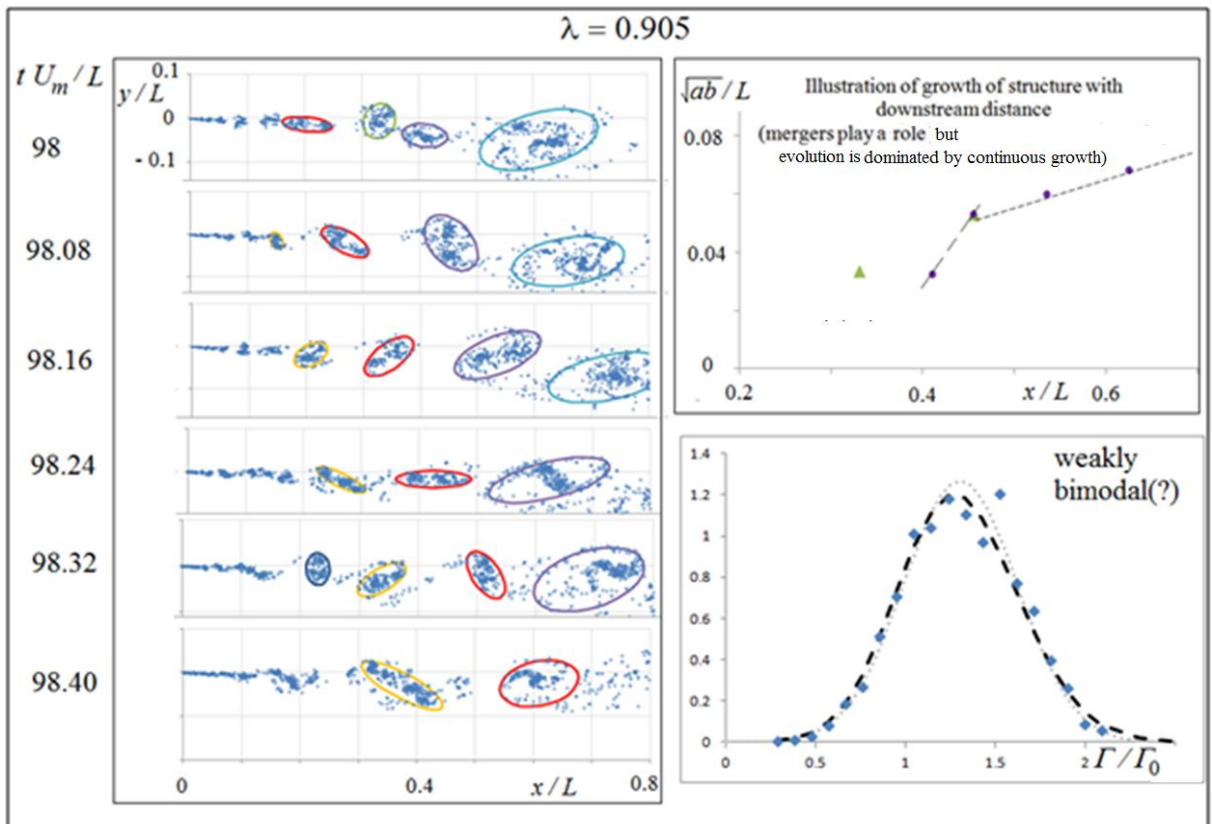

**Figure A3.** Analysis shown in Figs. 2, 3 and 4 for λ = 0.627, 1 repeated for an intermediate λ = 0.905. The results appear to suggest a mix of the two mechanisms, with domination of continuous structure-growth.